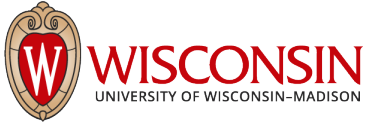

# Advancing technologies for high-resolution spatial and temporal measurements of macroscopic stellarator flows

A. M. Wright[1], D. J Den Hartog[2], B. Geiger[1], C. Lu[3], A. Wolfmeister[1], B. J. Faber[1]

[1]*Department of Nuclear Physics & Engineering Physics*
[2]*Department of Physics*
[3]*Department of Electrical & Computer Engineering*
University of Wisconsin – Madison

## 1. Introduction
Stellarators, together with tokamaks, represent the two mainstream approaches to realizing fusion energy via toroidal magnetic confinement of highly ionized gases – plasmas – at extremely high temperatures.

In stellarators, a magnetic field produced by external electromagnetic coils is sufficient to confine plasmas, but this magnetic field is non-axisymmetric, i.e., varies with toroidal angle, $\varphi$. By contrast, in tokamaks, a nominally axisymmetric magnetic field, which allows large toroidal flows to develop, is generated by external coils but must be supplemented by a magnetic field produced by a current driven within the plasma.

Consequently, the development of modern and next-generation stellarator configurations is centered on designing magnetic fields that produce desirable plasma confinement properties. Since stellarators admit magnetic fields that vary toroidally, the design space for the magnetic field and, consequently, the electromagnetic coils is greatly expanded compared to tokamaks.

In the US stellarator program, the dominant design strategy uses approximate symmetries of the magnetic field to improve plasma confinement properties, as predicted by theoretical analysis. These so-called *quasisymmetries* are a symmetry in the magnitude of the magnetic field in a particular coordinate system and, in theory, can lead to exact confinement of plasma particles, in an idealized limit.

Stellarators based on other design strategies also exist. For example, Wendelstein 7-X (W7-X) is based on a magnetic field that preserves particle confinement *on average*, again, in a theoretical, idealized limit.

## 2. Scientific motivation
Improving our understanding of how macroscopic flows impact equilibrium and dynamics in stellarators closes a significant knowledge gap and is necessary to advance the physics basis of stellarators as a fusion pilot plant (FPP) concept. This is especially critical for the US stellarator design program, in both the public and private sectors, where stellarator FPP concepts are based almost exclusively on leveraging quasisymmetries.

**There is strong experimental evidence of macroscopic flows in quasisymmetric stellarators:**
While flows have been observed to be strongly damped in W7-X, this is not true for stellarators in general. Indeed, the Helically Symmetric Experiment (HSX) – the world's only operating major quasisymmetric stellarator – has demonstrated quite the opposite. A series of experiments in the mid-2000s showed that substantial plasma flow can be generated in the direction of quasisymmetry (c.f. Figures 1 and 2). Moreover, events such as the core density collapse (CDC) events observed in Large Helical Device, lead to rapid (millisecond) and large-scale changes in the plasma profiles, necessarily generating macroscopic flows.

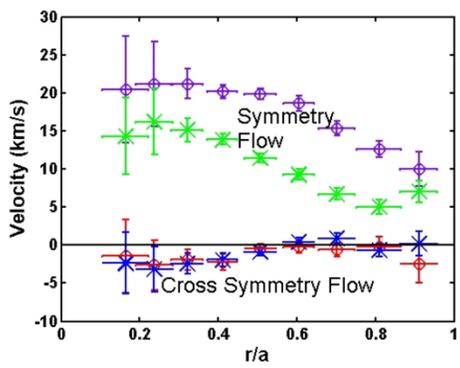

**Figure 1:** Reproduced from Fig. 3 of Ref. [1].

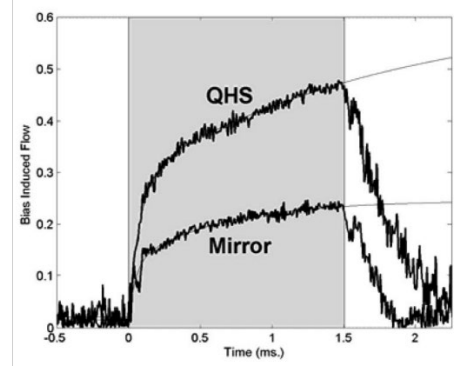

**Figure 2:** Reproduced from Fig. 2 of Ref. [2]

**Macroscopic flows can fundamentally alter the behavior of fusion plasmas:**
Intrinsically, fusion plasmas are extremely multi-scale and highly nonlinear. This means that macroscopic dynamics can affect the microscopic and vice versa, even when separated by 6 orders of magnitude, and that small changes can drive the system to an entirely different state. Thus, flows in a quasisymmetric stellarator needn't be fast to significantly impact performance. To illustrate this theoretically, consider a simple visco-resistive single-fluid magnetohydrodynamic model. In the conservation of momentum equation:

$$\rho \left(\frac{\partial \mathbf{v}}{\partial t} + \mathbf{v}\nabla \cdot \mathbf{v}\right) = \mu_0^{-1}(\nabla \times \mathbf{B}) \times \mathbf{B} - \nabla p + \nu \nabla^2 \mathbf{v}$$

There exists a nonlinearity in the fluid velocity (blue) and a nonlinearity in the magnetic field (red). The former also occurs in (neutral) fluid dynamics and is a convective nonlinearity associated with inertia of the fluid. The latter is due to the Lorentz force and arises only in plasmas, which consist of charged particles moving through an electromagnetic field. What's more, through the induction equation:

$$\frac{\partial \mathbf{B}}{\partial t} = \nabla \times (\mathbf{v} \times \mathbf{B}) + \eta \nabla^2 \mathbf{B}$$

that describes the evolution of the magnetic field in time, the fluid and magnetic field nonlinearities become coupled due to the Lorentz force (purple). This illustrates simply that a small change to flow and/or the magnetic field can be quickly amplified to produce a significant change in overall plasma performance.

**Fundamental assumptions are not always consistent with experiment:**
The assumption that a stellarator plasma exists in a static ($\mathbf{v} = 0$), steady ($\partial t \to 0$) equilibrium state at any instance in time, is pervasive in both theoretical and experimental settings. For example:
- Microphysics models that describe transport phenomena such as turbulence are frequently built assuming a static macroscopic (i.e., background) magnetic field.
- Equilibrium reconstruction is a key step in analyzing and interpreting experimental data.

However, the strongly nonlinear nature of stellarator plasmas (as illustrated above) means that the static, steady state assumption is not robustly valid, even if flows are small and quantities are varying slowly in time. In such a contradictory state, the correctness of results – both theoretical and inferred from experiment – may be questioned.

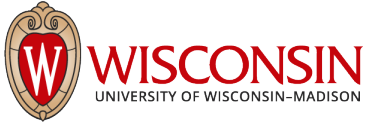

### 3. Timeliness
This issue becomes critical in the context of the forward extrapolations from which quasisymmetric stellarator FPP concepts are presently being designed and developed, especially in the US program.

The starting point of current, state-of-the-art stellarator optimization and design tools are equilibrium models which assume a plasma state that satisfies ideal MHD force balance:

$$\mathbf{J} \times \mathbf{B} = \nabla p$$

And neglects macroscopic plasma flows entirely ($\mathbf{v} = 0$). This includes optimization frameworks such as SIMSOPT and STELLOPT which are, or have been, widely used by the US stellarator community. However, from HSX measurements we *know* that significant flows are generated in quasisymmetric stellarators. To reconcile the inconsistency and improve the reliability of the projections on which future public and private investments (potentially worth billions of dollars) into stellarator FPP concepts are based, it is crucial to understand macroscopic plasma flows, especially in quasisymmetric stellarators.

More generally, strengthening the diagnostic capabilities necessary to advance the understanding of macroscopic flows in stellarators yields broad benefits beyond quasisymmetric stellarator FPPs. It has the potential to create synergies with existing international programs, such as efforts to measure flows in W7-X and LHD.

### 4. Opportunities
With this submission we invite the community to consider how the need for improving physics understanding of flows in advanced stellarators can serve as a catalyst for diagnostic and measurement innovation. This includes the identification of synergies with on-going technology development activities as well as opportunities to leverage new and emerging capabilities that may stem from hardware improvements and software, for example, application of AI/ML and workflow integration and management.

To help contextualize and ground the discussion, in the **Appendix** we highlight some diagnostic techniques where further innovation has the potential to deliver unique capabilities for high-resolution spatial and temporal measurements of macroscopic stellarator flows.

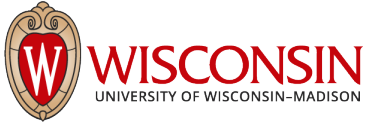

**Appendix. Selected examples of relevant diagnostic techniques**

Charge eXchange Recombination Spectroscopy: Charge eXchange Recombination Spectroscopy (CXRS) is based on the analysis of characteristic line radiation of impurity ions after receiving a bound electron via charge exchange reactions along a neutral beam [3]. Depending on the viewing geometry, the Doppler shift of the line radiation provides information on plasma flows in different directions. The typical time resolution of this type of measurement is 10 ms and, depending on the neutral beam width and geometry, can provide spatial resolution of few mm.

2D Coherence Imaging Spectroscopy: Coherence Imaging Spectroscopy (CIS) is a relatively novel technique which has been explored so far to provide ion flow velocities in the edge of fusion plasmas [4]. The diagnostic uses recent camera-based heterodyne polarization interferometers that allow for the detection of small Doppler-shifts of characteristic impurity line radiation, and hence provide 2D information on flow velocities. As an example, the flow pattern in the island divertor of W7-X could be studied using CIS [5].

X-ray imaging crystal spectrometer: X-ray Crystal Spectrometers have been deployed that provide wavelength resolved 1D spatial profile of line radiation emitted inside of fusion plasmas by highly charged impurity ions. Thanks to the 1D coverage, reconstruction of the measured impurity emission profile is possible so that spatially resolved Doppler shifts, and hence, plasma flows can be determined.

Turbulence propagation measurements: 2D fluctuation diagnostics such as Beam Emission Spectroscopy (BES), gas puff imagining (GPI) or Imagining ECE have been used to track turbulent eddies. As the Eddie-movement is impacted by plasma flows, information about plasma rotation is obtained. In addition, Doppler Reflectometry systems are commonly used that measure the propagation velocity of turbulent fluctuations based on the Doppler shift of back-scattered signals.

Finally, we remark that opportunities may present to leverage diagnostics that are already deployed to measure other plasma quantities such as density or electron temperature. One such diagnostic is the Fast Thomson Scattering system deployed on LHD. Innovations that enable the integration of multiple complementary techniques have the potential to yield unprecedented insight into the physics of macroscopic flows in stellarators.

Fast Thomson Scattering diagnostic: A Fast Thomson Scattering (TS) diagnostic measures electron temperature and density profiles at a rate of $\geq 1$ kHz. This is usually done in a burst mode, e.g. 20 kHz measurement rate for 5 ms. Dynamic changes such as those caused by ELMs, collapse events, and disruptions can be captured in this way. LHD currently fields a state-of-the-art Fast TS system, with measurement rates from 1 to 20 kHz, at 70 points across the diameter of the plasma. An example of the capability of this diagnostic is the measurement of the collapse of an electron ITB [6]. Planned next-generation Fast TS systems will be capable of measurement rates from 1 to 100 kHz. This will enable measurement of increased temporal detail during dynamic events and will begin to access kinetic phenomena such as temperature and density fluctuations responsible for turbulent transport.

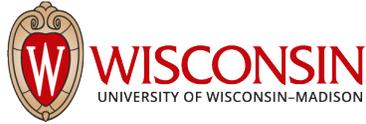